\documentclass[runningheads]{llncs}
\usepackage{algorithm}
\usepackage{amsmath,amssymb,amsfonts,bbm}
\usepackage{algpseudocode}
\usepackage{graphicx}
\usepackage{textcomp}
\usepackage{blkarray}
\usepackage{longtable}
\usepackage{rotating}
\usepackage[dvipsnames]{xcolor}
\usepackage[english]{babel}
\usepackage{cite}

\usepackage{multirow,tabularx}
\newcolumntype{Y}{>{\centering\arraybackslash}X}

\newcommand{\coo}{\ensuremath{\mathrm{CO_2}}}

\algnewcommand\algorithmicswitch{\textbf{switch}}
\algnewcommand\algorithmiccase{\textbf{case}}
\algdef{SE}[SWITCH]{Switch}{EndSwitch}[1]{\algorithmicswitch\ #1\ \algorithmicdo}{\algorithmicend\ \algorithmicswitch}%
\algdef{SE}[CASE]{Case}{EndCase}[1]{\algorithmiccase\ #1}{\algorithmicend\ \algorithmiccase}%
\algtext*{EndSwitch}%
\algtext*{EndCase}%

\begin{document}

\title{CAIS-DMA: A Decision-Making Assistant for Collaborative AI Systems}

\titlerunning{CAIS-DMA: A Decision-Making Assistant for Collaborative AI Systems}

\author{Diaeddin Rimawi\inst{1}\orcidID{0000-0003-3791-399X} \and
Antonio Liotta\inst{1}\orcidID{0000-0002-2773-4421} \and
Marco Todescato \inst{2}\orcidID{0000-0003-1449-5692} \and
Barbara Russo\inst{1}\orcidID{0000-0003-3737-9264}}
\authorrunning{D. Rimawi \textit{et al.}}

\institute{Free University of Bozen-Bolzano, Faculty of Engineering, Bolzano  39100, Italy 
\email{\{drimawi,antonio.liotta,barbara.russo\}@unibz.it}\\\and
Fraunhofer Italia, Bolzano 39100, Italy\\
\email{marco.todescato@fraunhofer.it}}

\maketitle              % typeset the header of the contribution

\begin{abstract}
% -----------
A Collaborative Artificial Intelligence System (CAIS) is a cyber-physical system that learns actions in collaboration with humans in a shared environment to achieve a common goal.
In particular, a CAIS is equipped with an AI model to support the decision-making process of this collaboration.
When an event degrades the performance of CAIS (i.e., a disruptive event), this decision-making process may be hampered or even stopped.
Thus, it is of paramount importance to monitor the learning of the AI model, and eventually support its decision-making process in such circumstances. 
This paper introduces a new methodology to automatically support the decision-making process in CAIS when the system experiences performance degradation after a disruptive event.
To this aim, we develop a framework that consists of three components: one manages or simulates CAIS's environment and disruptive events, the second automates the decision-making process, and the third provides a visual analysis of CAIS behavior. 
Overall, our framework automatically monitors the decision-making process, intervenes whenever a performance degradation occurs, and recommends the next action.
We demonstrate our framework by implementing an example with a real-world collaborative robot, where the framework recommends the next action that balances between minimizing the recovery time (i.e., resilience), and minimizing the energy adverse effects (i.e., greenness).

\keywords{Greenness \and Resilience \and Software Development Process \and Collaborative Artificial Intelligence System \and Cyber-Physical System } 
\end{abstract}

\section{Introduction}
\label{sec:intro}
\enlargethispage{\baselineskip}

A Cyber-Physical System (CPS) has heterogeneous hardware-software components that collaborate to deliver real-time services, \cite{rimawi_green_2022}. 
The complexity of CPS varies from one domain to the other.
A Collaborative Artificial Intelligence System (CAIS) is an example of a CPS that works together with humans to achieve a common goal, \cite{camilli_risk-driven_2021}. 
The core component of CAIS responsible for decision-making is its Artificial Intelligence (AI) model. 
The AI model is responsible for making decisions to control the collaboration between the system and the human. 
In general, AI model's training can be either from historical data (offline learning) or iterative during run-time (online learning), \cite{bonfanti_gresilience_2023}.
In CAIS context, the AI model learns from the human in an online learning mode.
Online learning can be affected by environmental changes (i.e., disruptive events) that may hamper the ability of the system to take real-time decisions.
For instance, disruptive events may affect the learning data, and thus, it may affect the AI model prediction accuracy and the reliability of the system, \cite{camilli_microservices_2022}.
Therefore, it is of paramount importance to provide CAISs with a recovery instrument that automatically supports the decision-making process in case of disruptive events.
The instrument needs to monitor the system performance, detect performance degradation, mitigate the cause through feasible recovery actions, and recover the system performance to an acceptable performance level, \cite{bonfanti_gresilience_2023, henry_generic_2012, colabianchi_discussing_2021}.

\enlargethispage{\baselineskip}
In this paper, we introduce our framework the \textit{Collaborative Artificial Intelligence System Decision-Making Assistant (CAIS-DMA)}, which automatically orchestrates the decision-making process between CAIS and humans when CAIS's performance degrades.
The framework is developed to be equipped as a CAIS component, monitors its performance under a disruptive event, and automatically intervenes when a performance degradation occurs.
The framework intervention aims to recover CAIS from performance degradation to an acceptable performance level. 
The recovery is achieved by supporting CAIS's AI model in restoring its accuracy as fast as possible, to ensure the real-time service delivery of CAIS.
To this aim, 
CAIS-DMA is equipped with three extendable components: i) Simulator, ii) Actuator, and iii) Monitoring component. 
The simulator simulates the run-time environment of CAIS's AI model and the human role in an online learning dataset. 
Then, it allows us to enforce the disruptive event effect on the dataset, and stream the data to the AI model in the expected structure.
On the other hand, the actuator monitors CAIS's performance and invokes the measurement mechanism to recommend the next action in case of performance degradation.
Finally, the monitoring component provides a toolbox for CAIS's managers to tune the framework components' configurations, and illustrates CAIS's behavior through a visual analysis representation.

Additionally, we demonstrate CAIS-DMA in a real-world demonstrator, in which we implement CAIS-DMA to assist a collaborative robot in recovering from performance degradation after a disruptive event occurs.
In this demonstration, CAIS-DMA will assist the robot in taking the next action that ensures fast recovery from the disruptive event (\textit{resilience}). 
However, this implies additional energy consumption, which increases the energy adverse effects and lowers CAIS's \textit{greenness}. 
Thus, we leverage our approach that balances the two properties. 
Our approach is \textit{GResilience} \cite{rimawi_green_2022, bonfanti_gresilience_2023}, a measurement mechanism to find the action that best trade-off between greenness and resilience. 
GResilience is equipped with two independent techniques: i) a weighted sum optimization model, and ii) a game theory model leveraging ``The Battle of Sexes" game.

Our major contribution in this paper can be summarized as follows:
\begin{enumerate}% restore
    \item We introduce CAIS-DMA our novel framework to assist CAIS's managers in simulating, actuating, and monitoring their systems. CAIS-DMA simulator supports creating a working environment with potential disruptive events, which allows testing CAIS's AI model responsible for the collaboration between the system and the human, without risking draining CAIS's resources. Additionally, the framework actuator automatically supports CAIS's decision-making by recommending the next action that achieves the selection criteria (through the selection mechanisms). Finally, CAIS-DMA provides a visual analysis monitoring component to monitor CAIS's performance.
    \item We design CAIS-DMA components to be extendable, where the developers can customize the framework components to represent different CAISs. The simulator can simulate different CAISs environments including the disruptive events they may be exposed to. Additionally, the actuator can be extended with new selection mechanisms and new recovery actions to recommend from. As for the monitoring component, it can tune the framework configurations to run different environmental settings.
    \item We show how CAIS-DMA can be equipped with CAIS, by demonstrating the development process with a real-world collaborative robot. In our application, we support action selection by recommending the action that best trade-off between greenness and resilience. Specifically, we wrap the GResilience \cite{rimawi_green_2022, bonfanti_gresilience_2023} as the measurement mechanism, to automatically support decision-making.
\end{enumerate}
\enlargethispage{\baselineskip}
The rest of this paper is structured as follows. In Sec.~\ref{sec:background} we provide a background about the performance states, resilience, greenness, and the GResilience approach. In Sec.~\ref{sec:frameworkArchitecture} we discuss CAIS-DMA architecture. In Sec.~\ref{sec:cais-dmadecisionmaking} we discuss how CAIS-DMA can automatically support the decision-making process in online-learning-based CAIS. In Sec.~\ref{sec:application} we demonstrate the development process of equipping CAIS-DMA with CAIS. In Sec.~\ref{sec:threatstovalidity} we discuss the paper threats to validity. In Sec.\ref{sec:relatedwork} we discuss the related work. Finally, in Sec.~\ref{sec:concandfuture} we state our conclusion and discuss our future work.

\section{Background}
\label{sec:background}
\begin{figure}[ht]
\includegraphics[width=\textwidth]{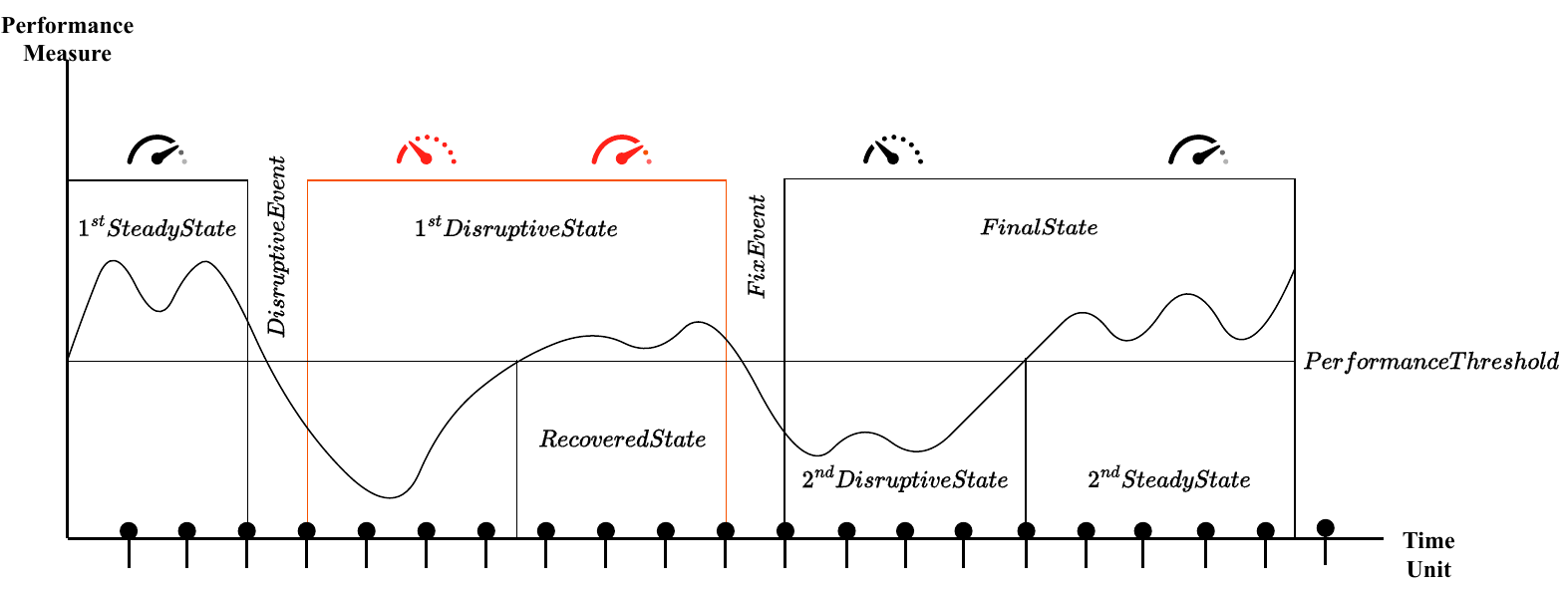}
\caption{CAIS Under Disruptive Event Performance States} \label{fig:resiliencestates}
\end{figure}

\enlargethispage{\baselineskip}
By principle, the online learning model eventually readjusts to the environmental changes after enough training, \cite{bonfanti_gresilience_2023}. 
Thus, CAIS's AI model, which is an online-learning-based model, learns based on accumulated data between normal and disruptive environmental settings. 
Hence, when fixing the disruption event CAIS's performance will face another performance degradation, due to the training data used while being under disruption.
Fig.~\ref{fig:resiliencestates}, shows the performance behavior of CAIS under disruption. 
The system starts with a \textit{Steady State}, in which the \textit{Performance Threshold} is defined as the lowest performance point. 
When the \textit{Disruptive Event} occurs it leads to performance degradation entering the \textit{Disruption State}. 
At this point, CAIS will try to adjust to the disruption, and with enough data from the \textit{human}, it manages to enter a \textit{Recovered State}. 
The system will continue living in disruption until a new environment change occurs (\textit{Fix Event}), then it enters the \textit{Final State}, starting with a second \textit{Disruptive State} due to the historical data, and then recover back to a second \textit{Steady State}.

CAIS-DMA aims to recommend actions based on decision criteria that are concerned with restoring the system performance from performance degradation to an acceptable performance level. 
Thus, it is important to define how to measure performance in order to understand what state the system is in.
% non-functional properties that we want to ensure during recovery.
In this paper, we focus on CAIS's AI model's ability for autonomous decision-making, and we consider the autonomous decision-making ratio in a windows of time as our performance measurement. 
As a result, we search for the action that both minimizes the time to restore the autonomous decision-making ratio from degradation to an acceptable level and minimizes the energy adverse effect. 
Thus, we consider two non-functional properties of CAIS: \textit{resilience} and \textit{greenness}.
The rest of this section discusses what they represent based on the GResilience measurement mechanism leveraged in our demonstration.

\subsection{Resilience}
\label{subsec:resilience}
\enlargethispage{\baselineskip}
% Make sure that we are not talking ourselves and not generalizing.
We consider resilience as a non-functional property that is concerned about recovering the system performance from a performance degradation to an acceptable performance level. 
Thus, we measure it by the action's \textit{Estimated Time (ET)}. 
Eq.~\eqref{eq:forecastingexponentialsmoothing}, uses the \textit{exponential smoothing} as an estimation technique to find the next iteration $i+1$ $ET$ for action $a$ ($ET_{a}^{i+1}$), where $(DT_{a}^{i} - ET_{a}^{i})$ is the error with the actual time from the previous iteration $i$, and $\alpha$ is the smoothing constant.
\begin{equation}
\label{eq:forecastingexponentialsmoothing}
   ET_{a}^{i+1} = ET_{a}^{i} + \alpha \cdot (DT_{a}^{i} - ET_{a}^{i})
\end{equation}

\subsection{Greenness}
\label{subsec:greenness}
\enlargethispage{\baselineskip}
System greenness is a non-functional property concerned with the efficient usage of energy while minimizing its adverse effects, \cite{kharchenko_concepts_2017}. 
One of the energy adverse effects we consider to measure CAIS greenness is the \coo{} footprint, which may increase when the CAIS's AI makes decisions to autonomously operate while still under disruption.
On the other hand, we can not drain human resources all the time by continuously moving to the learning mode. Thus, we constrain the human measurements with a maximum number of human iterations. 
To compute the system greenness we consider two variables: i)  \textit{Estimated \coo{} Footprint (ECF)} of an action, and ii) the human labor cost as \textit{Number of Human Interactions (NHI)} remaining for the action.
As for ET, the ECF is estimated using the exponential smoothing, Eq~\eqref{eq:carbonfootprintestimation}. $ECF_{a}$ is the ECF for action $a$ for an iteration $i$, $DCF_{a}$ is the actual carbon footprint of an iteration, and $\alpha$ is the smoothing constant.
Eq.~\eqref{eq:humanlaborcost} shows the $NHI_{a}$ which is the NHI for an iteration, $NHI_{max}$ is the maximum allowed NHI, and $NHI_{a}$ is the NHI required to complete action $a$.
\begin{equation}
\label{eq:carbonfootprintestimation}
   ECF_{a}^{i+1} = ECF_{a}^{i} + \alpha \cdot (DCF_{a}^{i} - ECF_{a}^{i})
\end{equation}
\begin{equation}
\label{eq:humanlaborcost}
   NHI_{a}^{i+1} = NHI_{max} - NHI_{a}^{i} - NHI_{a}
\end{equation}

\subsection{GResilience Measurement Mechanism}
\label{subsec:gresilience}
\begin{table}[t]
\caption{The GResilience Game General Payoff Matrix }
\centering
\scriptsize
\renewcommand{\arraystretch}{1.8}
\begin{longtable}[c]{l|l|c|c|c|}
     \multicolumn{2}{c}{} & \multicolumn{2}{c}{\pmb{$P_g$}} \\\cline{2-5}
    && \pmb{$a_1 (p)$} & \pmb{$a_2 (1-p)$} & \pmb{$P_r$  Expected Payoff} \\\cline{2-5}
     \multirow{2}{*}{\begin{sideways}\pmb{$P_r$}\end{sideways}}
     & \pmb{$a_1 (q)$} & $P^{2}_{r}(a_1), P^{2}_{g}(a_1)$ & $P^{1}_{r}(a_1), P^{1}_{g}(a_2)$ & $p P^{2}_{r}(a_1)  + (1 - p)  P^{1}_{r}(a_1)$\\\cline{2-5}
    & \pmb{$a_2 (1-q)$} & $P^{1}_{r}(a_2), P^{1}_{g}(a_1)$ & $P^{2}_{r}(a_2), P^{2}_{g}(a_2)$ & $p  P^{1}_{r}(a_2)  + (1 - p)  P^{2}_{r}(a_2)$ \\\cline{2-5}
    % \\\cline{2-5}
    & \pmb{$P_g$  Expected Payoff} & $q  P^{2}_{g}(a_1)  + (1 - q)  P^{1}_{g}(a_1)$ & $q  P^{1}_{g}(a_2)  + (1 - q)  P^{2}_{g}(a_2)$ & \\\cline{2-5}
\end{longtable}
\label{mat:payoffgeneral}
\end{table}
The GResilience (GR) approach \cite{rimawi_green_2022, bonfanti_gresilience_2023} provides CAIS with an automated instrument to support decision-making during disruption. GR aims to find the action that best trade-off between greenness and resilience. It solves the trading off problem by forming the problem into two independent mechanisms:
\begin{enumerate}
    \item Multi-Objective Optimization using the \textit{Weighted Sum Model} (GR-WSM): Where it combines both the resilience and the greenness measures into a single score per action. Then the model chooses the action with the highest score.
    Eq.\eqref{eq:globalscore}, shows the global score equation ($S()$) for the action $a$, where $w_{R}$ and $w_{G}$ are the weights of resilience and greenness respectively.
    $\epsilon$ is the confidence level of the AI model ($\epsilon \in [0, 1]$): the higher the value the more we trust the AI to continue operating. Thus, $\epsilon$  multiplies the inverse of the resilience measure, and $1 - \epsilon$ multiplies the summation of the greenness measures. Each resulting measure is then normalized ($N()$). Finally, we search for the action that maximizes $S(a)$.
    \begin{equation}
    \label{eq:globalscore}
            S(a) = w_{R} \cdot \epsilon \cdot N(ET^{-1}) + w_{G} \cdot (1 - \epsilon)\cdot\{N(NHI) + N(ECF^{-1})\}
    \end{equation}
    \item Game Theory by leveraging ``The Battle of Sexes" game into building \textit{The GResilience Game} (GRG): A collaborative game where each of the players has a preferred option, while they share a common goal (i.e., recovering the system). The game treats resilience and greenness properties as two game players ($P_{r}$ the resilience player and $P_{g}$ the greenness player), and each of the players has an independent way to measure its payoff. Same as ``The Battle of Sexes" the GRG has two Pure Strategies Nash Equilibria (PSNE), in which the two players agree on the same action. GRG may have another Mixed Strategy Nash Equilibrium (MSNE) based on the probability of each player's action.
    Eq.~\eqref{eq:resiliencepayoff} shows the expressions to find the $P_{r}$ and $P_{g}$ payoffs,
    where $\alpha$ is the matching factor that is a smaller value in case the players land on different actions and a larger value in case they match.
    Table~\ref{mat:payoffgeneral}, shows two PSNEs where both players choose the same action and a possible MSNE based on the probability of each player's action, \cite{rimawi_green_2022, stowe_cheating_2010}.
    In the MSNE, $P_r$ chooses $a_1$ with probability $q$ and $a_2$ with probability $1-q$, while $P_g$ chooses $a_1$ with probability $p$ and $a_2$ with probability $1-p$, which results to the \textit{expected payoff} described in Table~\ref{mat:payoffgeneral}. Thus, to find the probability $q$ (resp. $p$) with MSNE, we equal the expected payoffs of $P_g$ (resp. $P_r$) for $a_1$ and $a_2$ and solve the resulting equation for $q$ (resp. $p$).
    \begin{equation}
    \label{eq:resiliencepayoff}
       P^{\alpha}_{r}(a) = \epsilon \cdot \alpha \cdot ET^{-1}, \,\,
       P^{\alpha}_{g}(a) = (1 - \epsilon) \cdot \alpha \cdot NHI^{-1} \cdot ECF^{-1}
    \end{equation}
\end{enumerate}

\section{Framework Architecture - CAIS-DMA}
\label{sec:frameworkArchitecture}
\begin{figure}[t]
\includegraphics[width=\textwidth]{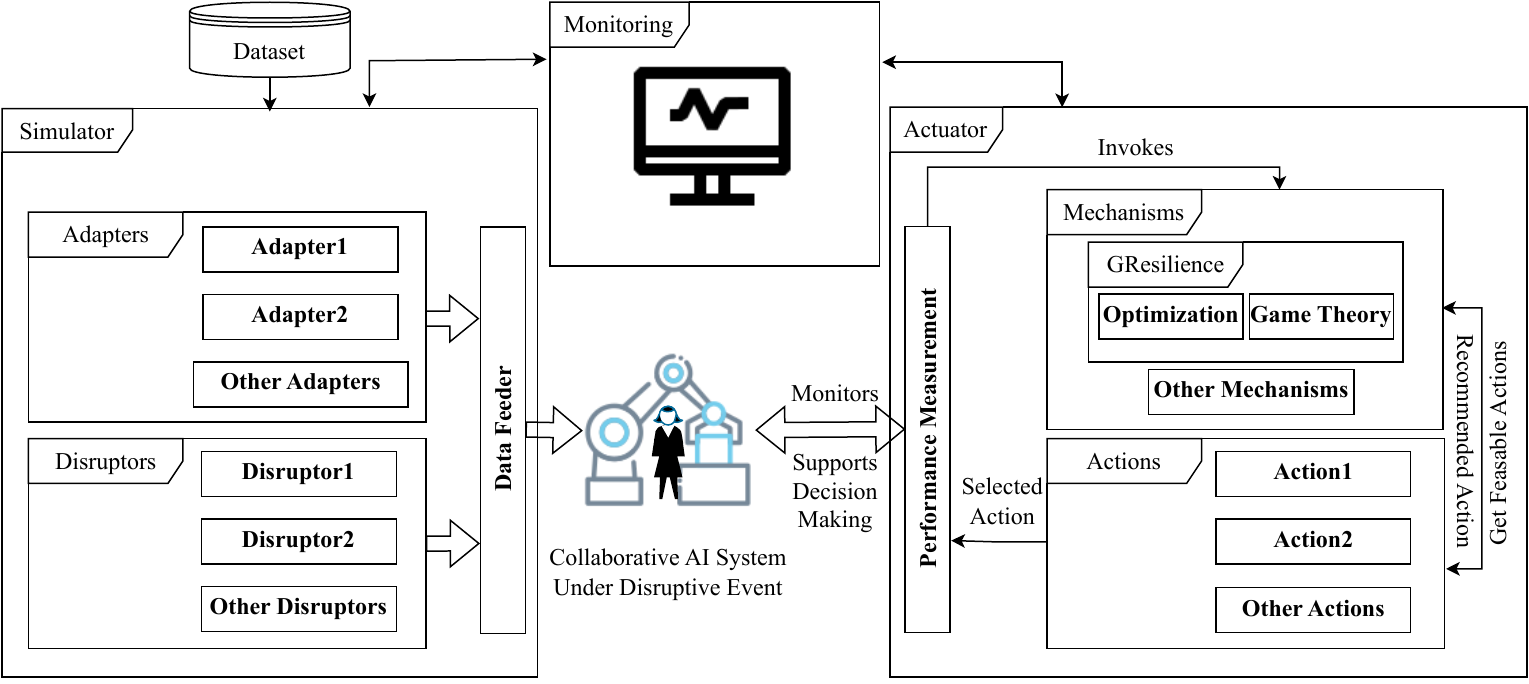}
\caption{CAIS-DMA Architecture} \label{fig:caisdma}
\end{figure}

\enlargethispage{\baselineskip}
The CAIS-DMA operates as an assistant to support CAIS's AI model responsible for the collaboration actions between the system and the human during the first disruptive state (Fig.~\ref{fig:resiliencestates}). 
It aims to assist the AI model until it reaches the recovered state.
The framework consists of three components: i) a data simulator, ii) a decision-making actuator, and iii) a monitoring component. Fig.~\ref{fig:caisdma}, shows the three components interacting with the CAIS under the disruptive event. The rest of the section will discuss each of CAIS-DMA's components.

\subsection{Simulator}
\label{subsec:simulator}

The goal of the simulator component is to simulate the learning data of CAIS's AI model, editing the data to represent a specific disruptive event effect and structure it as expected by the AI model. 
To this aim, the simulator consists of two packages: 
\begin{enumerate}
    \item \textit{Adapters}: An extendable package that contains the adaptation classes. 
    These classes prepare the dataset and restructure it to adapt the AI model's expected input.
    Moreover, it adds a data field for human representation. The human field is assumed to be the ground truth of the specific data instance.
    This field is important to simulate the human action for the specific data instance.
    \item \textit{Disruptors}: The disruptors package is an extendable package, where it allows multiple disruptors. Each disruptor aims to simulate a disruptive event effect on the data, and it is important to note that the disruptors affect only the data and not the system itself.
\end{enumerate}  

After data preparation, the simulator streams the data instances to the AI model using the \textit{Data Feeder}. 
The data feeder is responsible for sorting the data and streaming them over to the AI model. 
It streams the data in three states: i) Steady State, where it streams the data without disruptions, ii) Disruption State, where it streams the data with the disruptive effect, and iii) Final State, where it streams again the data without disruptions (to simulate the disruptive event fix). 
By default, the data are split into thirds, unless defined otherwise in the framework configurations.

\subsection{Actuator}
\label{subsec:actuator}
\enlargethispage{\baselineskip}
The actuator component has two main functionalities: i) Monitoring the AI model performance, and ii) Supporting the AI model decision-making. The following summarizes these functionalities:
\begin{enumerate}
    \item \textit{Performance Monitor.} This package monitors the AI model's decisions and measures its performance based on the autonomous decisions made for a window of time. 
    Algo.~\ref{algo:performancemeasure}, shows the algorithm we use to measure CAIS's AI model performance. The algorithm first initializes the variables, measures the performance, invokes CAIS-DMA decision mechanisms, and executes the final decision (recommended action).
    Where $ADR$ is the autonomous decision ratio, $ADRT$ is the $ADR$ threshold that defines what an acceptable performance level, $D$ is the AI's decision (chosen/recommended action), $\epsilon$ is the AI model confidence level towards the decision $D$, $W$ is the time window size, $Q$ is a queue that stores the last $W$ decisions.
    \item \textit{Support Decision-Making.} In case of performance degradation the performance monitor invokes a decision-making mechanism to recommend the action to be executed. 
    All feasible actions are defined in the \textit{actions package} including their properties, for example, their execution time. 
    The decision-making mechanisms represent different techniques for decision-making, such as the GResilience measurement mechanism which recommends the action that best trade-off between greenness and resilience. Other measurement mechanisms can be defined in this package to support the different nature of CAIS under test.
\end{enumerate}
\enlargethispage{\baselineskip}
\begin{algorithm}[htb!]
\caption{Performance Measurement Algorithm}
\label{algo:performancemeasure}
\scriptsize
\begin{algorithmic}[1]

\State $W \gets Window Size$  \Comment{\textit{Variables Initialization}}
% \State $M \gets Minimum Probability$
\State $ADRT \gets ADR Threshold$
\State $Q \gets Queue()$
\For{$i = 0 \to W$}  
    \State $Q.enqueue(0)$
\EndFor
\While{True} \Comment{\textit{Keep Monitoring The CAIS's AI Decisions}}
    \State $D, \epsilon \gets ReadDecisionAndProbability()$  \Comment{\textit{D: AI Decision, $\epsilon$: Confident Level}}
    \State $Q.dequeue()$ \Comment{\textit{Update the Queue}}
    \Switch{$D$}
        \Case{``Autonomous"}: $Q.enqueue(1)$
        \EndCase
        \Case{``Human"}: $Q.enqueue(0)$
        \EndCase
    \EndSwitch
    \State $ADR \gets Q.sum() / W$ \Comment{\textit{Calculate ADR}}
    \If{$ADR < ADRT$} \Comment{\textit{Check for Performance Degradation}}
        \State $D \gets InvokeDecisionMechanism(D, \epsilon)$ \Comment{\textit{Recommend Decision $D$}}
    \EndIf
    \State $ExecuteDecision(D)$ \Comment{\textit{Proceed with the Decision $D$}}
\EndWhile
\end{algorithmic}
\end{algorithm}
\vspace{-15px}
\subsection{Monitoring Component}
\label{subsec:monitor}
\enlargethispage{\baselineskip}
The monitoring component is a web-based application, which provides a toolbox for the framework users to visually analyze CAIS performance during run-time, and tunes the framework variables to run experiments over CAIS under test. 
Through visual analysis, we can monitor the performance anomalies caused by the disruptive event(s), where for each experiment it plots 
the performance behavior per each window of time.
While through the experiment tuning, we can customize an experiment by setting the experiment configurations, like the \textit{number of iterations}, the \textit{dataset}, the \textit{adapters} to use, the \textit{disruptors}, what \textit{decision-making mechanism} to apply, and the \textit{actions set} to recommend from.

\section{Support Decision Making Process with CAIS-DMA}
\label{sec:cais-dmadecisionmaking}
\begin{figure}[t]
\includegraphics[width=\textwidth]{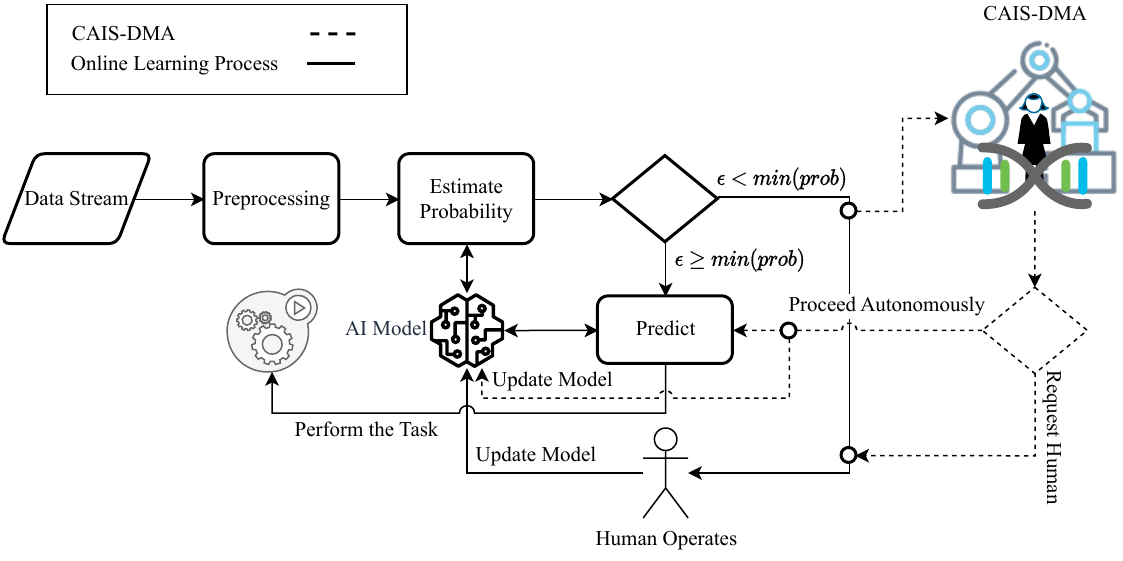}
\caption{Online Learning with CAIS-DMA} \label{fig:onlinelearning}
\end{figure}

\enlargethispage{\baselineskip}
The AI model of CAIS is responsible for controlling the collaboration between the human and the system.
It decides whether to run autonomous actions by the system or ask the human to perform the action and update the AI model with the new learning.
Fig.~\ref{fig:onlinelearning}, shows the online learning flow diagram of CAIS, where it starts by receiving a new data stream, preprocessing the data, and then estimating the prediction probability (i.e., the confidence level $\epsilon$, where $\epsilon \in [0, 1]$) using the AI model to perform a specific task.
If the prediction probability is more than the predefined minimum probability ($min(prob)$), the decision will be to perform autonomous actions, through predicting and performing the task by the system itself.
Otherwise, it asks the human to perform the task, by entering into a learning mode and updating the AI model with the new data.

To support the decision-making process, CAIS-DMA monitors the estimated probability of the online learning process to measure CAIS's performance. This measurement helps CAIS-DMA to automatically detect disruptive events that lead to performance degradation.
If performance degradation is detected, CAIS-DMA calls the decision measurement mechanism, which, in turn, considers $\epsilon$ with the feasible actions to measure the action that best trade-off between the predefined non-functional properties. 
In online learning, CAIS-DAM selects between requesting human intervention or proceeding autonomously.
Requesting human intervention simply means continuing with the original flow.
While proceeding autonomously, means moving forward with CAIS tasks by allowing predictions.

Finally, CAIS managers have the flexibility to update the AI model with the new prediction or not.
If they decide to update the AI model, this reduces CAIS dependability on CAIS-DMA, allowing the system to live under disruption without the framework support, however, it causes a higher probability to face another disruptive state after fixing the disruptive event, due to the accumulative learning.
While, if they avoid updating the AI model, this means discarding learning from disruptive data, which reduces the chances of being disrupted again after fixing the disruptive event, but increases the dependability over CAIS-DMA during the disruptive state.

\section{First Application and Evaluation}
\label{sec:application}
\enlargethispage{\baselineskip}
To understand if CAIS-DMA supports the decision-making of CAIS to recover from performance degradation caused by disruptive events, we aim to answer the following research questions:
\begin{itemize}
    \item \textbf{RQ1}: \textit{What is the software development process of CAIS-DMA to automatically support the decision-making of CAIS?} To answer this question, we will demonstrate development process decisions to successfully design an application of CAIS-DMA, using both its simulator and actuator to automatically support the decision-making of CAIS.
    \item \textbf{RQ2}: \textit{What are the extendable components of CAIS-DMA that allow wider support for decision-making?} To answer this question, we will consider CAIS-DMA architecture with the online learning process to reflect on a real-world demonstrator, showing the framework components to be extended in order to complete a full application.
\end{itemize}

\subsection{Application Context - CORAL}
\label{subsec:coral}
\begin{figure}[t]
\includegraphics[width=\textwidth]{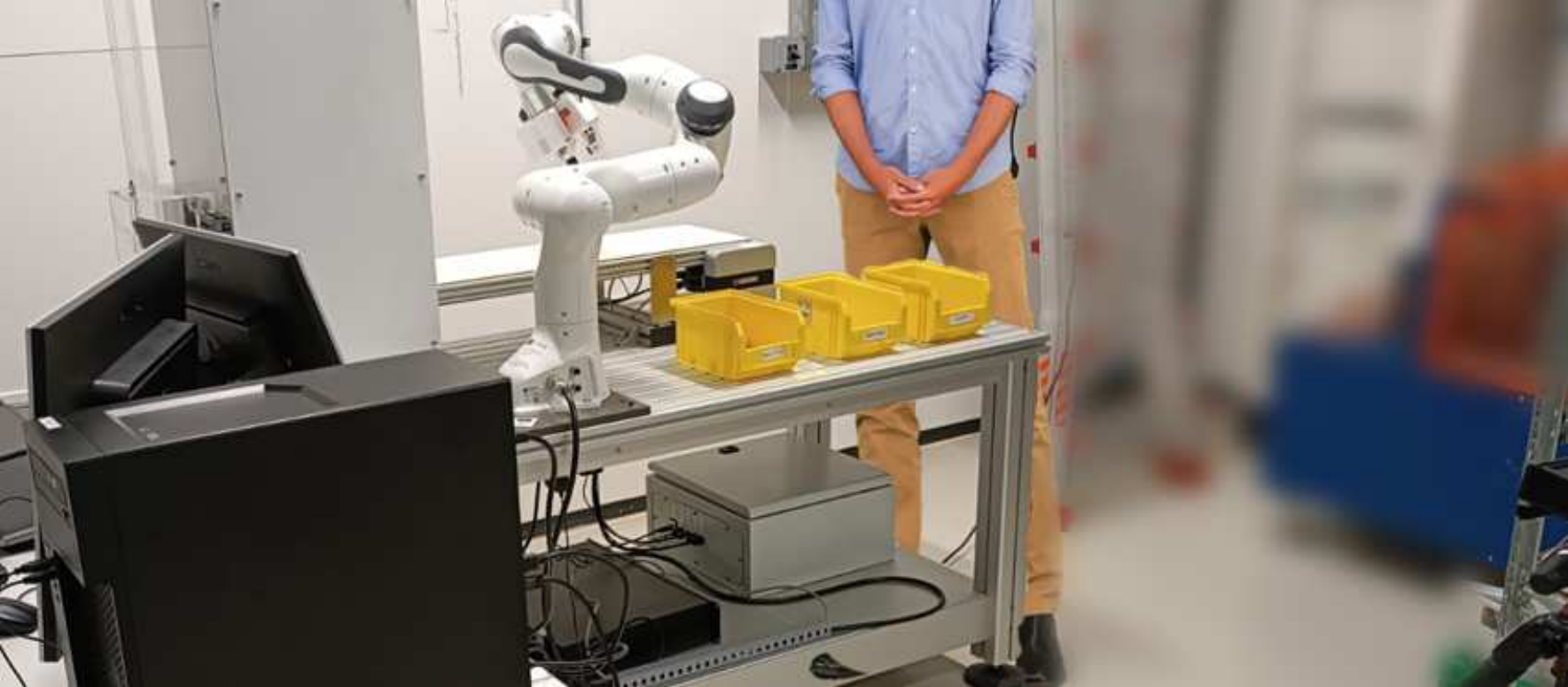}
\caption{Collaborative Robot Learning from Demonstrations - CORAL} 
\label{fig:coral}
\end{figure}

\enlargethispage{\baselineskip}
Our demonstrator is a collaborative robot named ``CORAL\footnote{CORAL is developed by Fraunhofer Italia Research in the context of ARENA Lab}". 
Fig.~\ref{fig:coral}, shows CORAL, which consists of a robotic arm installed above a conveyor belt that transfers objects of multiple colors. 
The robot detects the objects using an RGB camera installed on top of the conveyor belt. 
The detected object is streamed to an \textit{online learning model} to be classified based on its color. 
The classifier monitors human movements by tracking the human skeleton. The human movement helps the classifier to label the object class with the target box.

\subsection{Equipping CORAL with CAIS-DMA}
\label{subsec:coralwithcais-dma}
\begin{figure}[t]
\includegraphics[width=\textwidth]{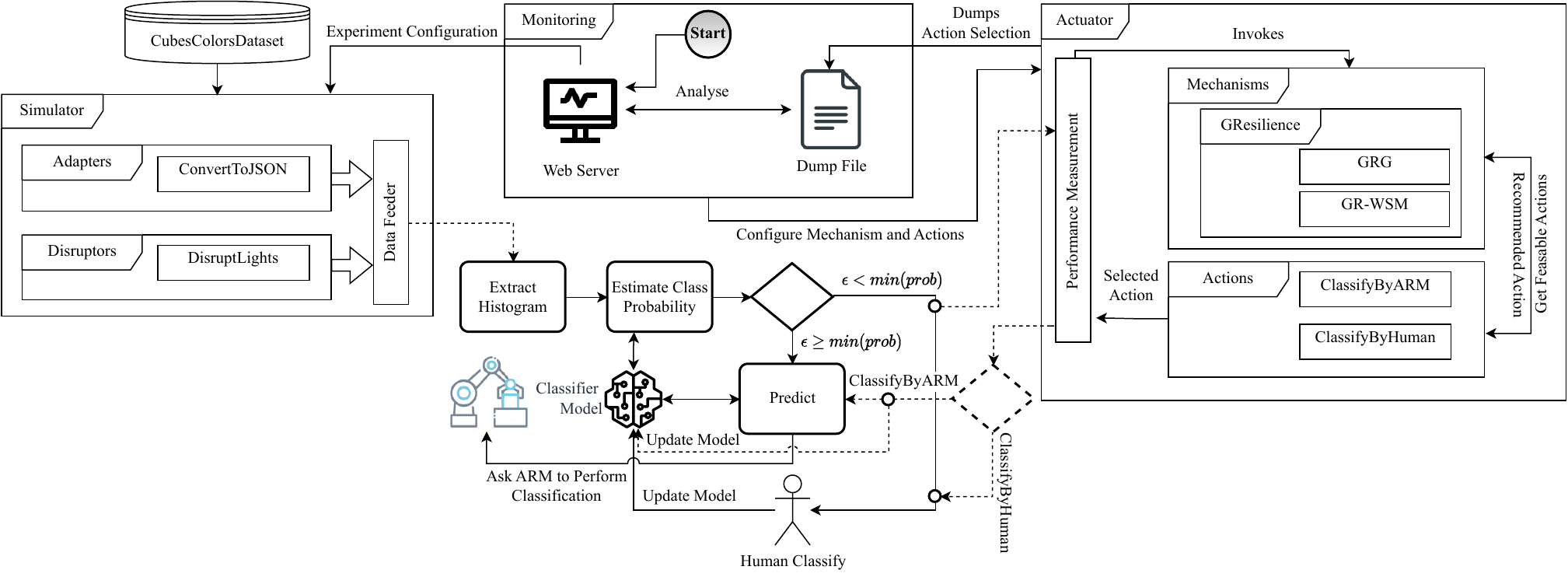}
\caption{Equipping CORAL with CAIS-DMA, Simulate Cubes Colors Dataset, Support Decision Making using GResilience Mechanism, and Analyze Performance}
\label{fig:coralarchwithcais-dma}
\end{figure}

\enlargethispage{\baselineskip}
CAIS-DMA framework is built using \textit{Python3}, and its available on \textit{GitHub}\footnote{\url{https://github.com/dmrimawi/CAIS-DMA}}.
Fig.~\ref{fig:coralarchwithcais-dma}, shows CORAL online learning associated with CAIS-DMA. The online learning of CORAL starts when a new object is detected, the detected object is streamed to the preprocessing step. The preprocessor extracts the object histogram and passes it to estimate the class probability. 
If $\epsilon \ge min(prob)$ ($min(prob) = 0.4$ in the case of CORAL) the classifier predicts the object's box and asks the robotic arm to pick the object to the predicted box.
Otherwise, the classifier notifies the human to classify the object and update the model with the new labeled object.
To the aim of building a successful application using CAIS-DMA, we illustrate the software development process of building a green resilient CORAL. The process milestones are: i) Defining the performance measurements, ii) Understanding the dataset structure the AI model expects, iii) Listing the disruptive events that may lead to performance degradation, iv) Creating a set of all feasible recovery actions, v) Defining the non-functional properties the actions have to balance,  vi) Setting the decision-making measurement mechanisms, vii) Storing the collected decisions, and finally, viii) Updating the AI model with the decision.
\enlargethispage{\baselineskip}

The rest of this section summarizes the decisions with respect to CORAL:
\enlargethispage{\baselineskip}

\noindent\textbf{Performance Measurements.} In our application, we monitor the estimated classification probability from CORAL's AI model, to count the number of autonomous operations against the number of human operations.
Then we compute the autonomous classification ratio in a window of time ($W$-consecutive operations).
CAIS-DMA supports running experiments with a range of values for $W$, allowing CAIS managers to tune the value to what suits their systems. 

\noindent\textbf{Cubes Colors Dataset.} As illustrated CORAL's learning starts when a new object is detected over the conveyor belt. The detected object is streamed into a JSON-structured instance. This instance is then processed by the learner preprocessing step. 
CAIS-DMA simulator's goal is to simulate the input data (i.e., the cube images in this case), thus, we created the \textit{Cubes Colors Dataset}, a set of cubes images taken from the RGB camera installed above the conveyor belt of CORAL, to have a real-world dataset from the application itself.
Finally, we create an adapter that prepares the images to be streamed in the same structure CORAL's AI model expects.

\noindent\textbf{Disruptive Event.} One of the disruptive events that lead to performance degradation for online learners depending on computer vision to produce the learning data, is disrupting the vision itself. 
Thus, in this application, we simulate the environmental effect of losing the lights above the conveyor belt, which, in turn, leads to darker images. 
As we are using the simulator to produce images' darkness, this gives us the flexibility to simulate the disruptive event effect from dimming the lights to losing the lights completely.
The effect of darkness hampers online learning as it produces different histograms than the training data before the disruptive event, which disrupts the decision-making and eventually leads to performance degradation (low autonomous classification ratio). 

\noindent\textbf{Recovery Actions.} The online learning classifier of CORAL is moving between two states: i) the learning state, where it asks the human to classify the object, and ii) the operating state, where it predicts the cube box and places it using the arm. 
Both of these states are considered as the decision actions, and the iterative execution of these actions will eventually lead to recovery from the performance degradation.
However, choosing which action to execute will depend on the action that best trade-off between the non-functional properties selected in the measurement mechanism.
To summarize, in our application we consider two feasible recovery actions from the light disruption: \textit{action1}) ask the arm to classify the cube, and \textit{action2}) learn from the human classification of the cube. 
\enlargethispage{\baselineskip}

\noindent\textbf{Greenness and Resilience.} The main goal of this application is to restore CAIS performance from degradation to an acceptable performance state. In other words, ensure the \textit{resilience} property of CAIS.
Although choosing the \textit{fastest action} all the time may restore CAIS performance faster, it leads to higher energy consumption, which increases the energy adverse effects.
Increasing the energy adverse effects reduces CAIS \textit{greenness}, thus, it is important to find the action that balances CAIS's greenness and resilience.
Hence we use ECF and NHI to measure greenness and ET to measure resilience.

\noindent\textbf{GResilience Measurement Mechanism.} The GResilience approach aims to recommend the next action that best trade-off between greenness and resilience to restore CAIS's performance to an acceptable performance state.
The approach defines two techniques to find the next action, either by recommending the action with the highest combined score of resilience and greenness through optimization, or by considering the two properties as game theory players in a collaborative game of common goal (Sec.~\ref{sec:background}).

\noindent\textbf{Storing the Results.} To be able to revisit the results, analyze them, and support CAIS's managers in making decisions based on these results,  it is important to structure these results in a readable and productive way.
In our application we dump the decisions into a Comma-Separated Values (CSV) structure, in which we record timestamped data that contains the selected action, the measurements (ET, ECF, and NHI), and if this decision is made autonomously by CORAL, or it is overridden through CAIS-DMA.

\noindent\textbf{Update the AI Model.} Finally, CAIS managers need to decide whether to update the AI model with CAIS-DMA results or not. Based on our observation, updating the AI model leads to faster recovery during the disruption state, in other words, CAIS learns to \textit{live in the darkness}. However, this leads to a second disruptive state after fixing the disruptive event.
On the other hand, not updating the AI model, leads to higher dependency on CAIS-DMA during a disruptive state, making it difficult to learn to live in the darkness. 
While, after fixing the disruptive event it is faster to return to the steady state again.
We recommend running both options under simulation and taking the decision of updating the AI model or not based on the CAIS under test.

\subsection{Takeaways}
\label{subsec:takeaways}
\enlargethispage{\baselineskip}
In this application, we equipped our real-world demonstrator (CORAL) with CAIS-DMA. 
The implementation process of our application helps address our research questions as follows:

\noindent\textbf{Answer to RQ1.} The answer to this research question aims to demonstrate the development process for CAISs managers. 
In particular, the managers need to define the measures that are relevant to observing CAIS's performance, and what are the events that may lead to performance degradation. 
Then, they have to address the feasible recovery actions, and the recommendation criteria for the next action.
In this application, we show an application of the development process decisions based on CORAL, where we used CAIS-DMA to simulate CORAL's working environment and the effect of losing the light on the vision sensor of CORAL. 
Then we implemented CAIS-DMA actuator to recover CORAL by balancing between greenness and resilience, and finally, we monitored CORAL behavior to understand the performance behavior during the different states.

\noindent\textbf{Answer to RQ2.} This question provides CAIS's developers with the main extendable packages to consider during implementation.
Starting with the simulator, in our application, we created a dataset of images to simulate CORAL's environment, however, the developers can customize the simulator to simulate their own dataset. They just need to provide the adapter class and the disruptor to feed the data to CAIS under test.
Additionally, the mechanisms and the recovery action packages in the actuator component are also extendable to different measurement mechanisms and recovery actions.
Finally, CAIS-DMA is full of utilities and toolboxes to automate running experiments, visual analysis, and sorting the dataset instances.

\section{Threats to Validity}
\label{sec:threatstovalidity}
\enlargethispage{\baselineskip}
Wieringa \textit{et al.} \cite{wieringa_six_2015} define validity as the support degree to a fallible inference. Our paper introduces an architectural explanation of CAIS-DMA, to automatically support decision-making of CAIS during disruptive state. 
The threats to the \textit{internal validity} in our paper are represented in the degree of support to the architectural explanation, where we have selected the GResilience approach throughout the explanation. In this respect, we did not consider the human cost and energy in the overall measurement, which left to future work.

Threats to \textit{external validity} are related to the degree of support for the generalization of the architectural explanation to a theoretical population. Which in our case is related to the application implementation that has been applied to the specific domain of CAIS, and CORAL robot. However, the framework itself is designed to be general.

Finally, \textit{conclusion validity}, which represents the degree of support of a statistical inference from a sample to a study population. It is important to state that the nature of our framework is mainly exploratory, and generalization beyond the CORAL application is needed for consolidating our claims. 
Thus,  we plan to run more experiments with another real-world demonstrator we have in-house and with simulators.

\section{Related Work}
\label{sec:relatedwork}
\enlargethispage{\baselineskip}
We have reviewed existing literature according to three lines of research: i) The non-functional properties to balance, specifically resilience and greenness, ii) The trading off techniques using optimization and game theory, iii) The decision-making support framework in CAIS context. In the following, we illustrate a brief overview of them.

\noindent\textbf{Resilience and Greenness.} 
Methods and frameworks aim to build a resilient system have been discussed extensively by the literature, like using a multi-agent model by Januário \textit{et al.} \cite{januario_distributed_2019}, tri-optimization model by Liu \textit{et al.} \cite{liu_distributionally_2022}, and deep learning model by Zarandi \textit{et al.} \cite{zarandi_detection_2020}, to mention a few.
The major goal of these methods is to restore the system performance from degradation caused by disruptive events to acceptable performance.
The disruptive events are different depending on the system itself, for example, the disruptive event can be a security vulnerability of the system \cite{zarandi_detection_2020, liu_distributionally_2022}, a defect in the software or hardware parts \cite{januario_distributed_2019}, or caused by humans \cite{bonfanti_gresilience_2023}.
In this paper, we add an additional requirement to restore the system's performance. We are interested in restoring the performance while monitoring and controlling the energy adverse effects, which is how Kharchenko \textit{et al.} \cite{kharchenko_concepts_2017} define greenness. Studies have discussed greenness as a default result of building a resilient system, such as, Pandey \textit{et al.} \cite{pandey_greentpu_2020}. Other studies 
seek to find a trading-off between greenness and resilience, \cite{mohammed_eco-gresilient_2018, bonfanti_gresilience_2023}.

\noindent\textbf{Optimization and Game Theory.}  In the greenness and resilience context, Mohammed \textit{et al.} \cite{mohammed_eco-gresilient_2018}, propose a solution to optimize supply chain network distribution using the eco-gresilient model, which trades off between three objectives, specifically economical, green, and resilient. They used the proposed solution to find the best number of facilities in the supply network section. 
Game theory is a decision-making process with multiple actors. For instance, Xu \textit{et al.} \cite{xu_user_2018}, defined a collaborative game to support the decision process for a recommendation system for users' satisfaction.
To the best of our knowledge, using game theory to trade off non-functional properties is a novel idea that we have sketched in our previous work, \cite{bonfanti_gresilience_2023}. 
In this current paper, we have built a novel framework that operationalizes our initial idea and we have exemplified it to CORAL. We have further worked on a case study on CORAL, which is now under submission, \cite{rimawi_2023_gresilience}.

\noindent\textbf{Decision-Making Assistant.} Various studies handle the automatic support of decision-making frameworks in CAIS context, by considering humans as the ground truth of the system's actions. They build specific knowledge about human actions and then use the knowledge to support decision-making by inferring human actions.
For example, Chen \textit{et al.} \cite{chen_trust-aware_2020}, build a computation model to assess the human trust in CAIS autonomous actions, and then it uses this assessment to automatically support decision-making with the action that maximizes the trust value. 
Other studies predict human actions using AI-based techniques, such as Ghadirzadeh \textit{et al.} \cite{ghadirzadeh_human-centered_2020}, using deep reinforcement learning, and Quintas \textit{et al.} \cite{quintas_toward_2018}, who built an AI-agent that monitors human actions and generates descriptive scenarios to automatically support CAIS's decisions.
As there are several frameworks to support decision-making, to our knowledge none of these frameworks provide extendable components backed with the toolbox and utilities needed, to first, simulate the system environment, second, automatically support decision-making through different decision-making mechanisms, and third visually analyze different experimental configurations, which make our framework (CAIS-DMA) a novel framework in that sense.

\section{Conclusion and Future Work}
\label{sec:concandfuture}
\enlargethispage{\baselineskip}
\noindent\textbf{Conclusion.}
% Add about the option we update the module 
In this paper, we introduce our novel extendable framework \textit{CAIS-DMA} to automatically support CAIS decision-making in an online learning process. CAIS-DMA aims to deal with different learning situations, for this reason, it contains a simulator, actuator, and monitoring component.
The framework simulates CAIS environment and represents the potential disruptive events that face the specific CAIS.
It monitors CAIS's performance to detect any performance degradation to automatically support the decision-making to recommend the actions that help in restoring the performance to an acceptable level. 
CAIS-DMA monitors the decision made by CAIS's AI model and overrides the selected action, by the action that best trade-off between the non-functional properties defined through measurement mechanisms.
Additionally, CAIS-DMA supports running different experiments on CAIS's AI model through different experiment configurations and provides a visual analysis of the performance behavior.
Finally, we demonstrate the framework through a real-world demonstrator, showing the implementation steps and the framework's extendable components.
\enlargethispage{\baselineskip}

\noindent\textbf{Future Work.}
\enlargethispage{\baselineskip}
This framework allows us to conduct a wider range of experiments with simulated data. 
Thus, we plan to run an experiment to compare the results from the simulation and the real-world case. 
Secondly, we plan to extend the framework in order to explore other non-functional properties such as safety. For instance, we can create a disruptive event that simulates an attack that alters the safe distance between the human and the robotic arm. 
In this case, we will study the actions that trade-off between human safety and system performance.
Moreover, we plan to use the measurement mechanism results as reinforcement learning of the system that rewards recommended decisions. 

% \begin{equation}
% \label{eq:forecastingexponentialsmoothing}
% \begin{split}
%    Q(S_t, A_t) & \longleftarrow Q(S_t, A_t) + \alpha [\{w_R \cdot \epsilon \cdot N(ET^{-1}) \\
%    & + w_G \cdot (1-\epsilon) \cdot (N(NHI) + N(ECF^{-1})) + \beta\} \\
%    & + \gamma \cdot max_aQ(S_{t+1}, a) - Q(S_t, A_t)]
% \end{split}
% \end{equation}

% \begin{equation}
% \label{eq:forecastingexponentialsmoothing}
% \begin{split}
%    R_{t+1} & \longleftarrow w_R \cdot \epsilon \cdot N(ET_{a_{t+1}}^{-1}) \\
%    & + w_G \cdot (1-\epsilon) \cdot [N(NHI_{a_{t+1}}) + N(ECF_{a_{t+1}}^{-1})] \\
%    & + \beta
% \end{split}
% \end{equation}

% \[
%     \beta = 
% \begin{cases}
%     1,& \textbf{if } S' \in [Red, Green, Blue]\\
%     0,& \textbf{if } S' \in [Black]
% \end{cases}
% \]

\bibliographystyle{splncs04}
\bibliography{references}

\end{document}